\documentclass[journal]{IEEEtran}

\ifCLASSINFOpdf
\else
   \usepackage[dvips]{graphicx}
\fi
\usepackage{url}

\usepackage{pbox}
\usepackage{tikz}
\usepackage{pgfgantt}
\usepackage{subfigure}
\usepackage{gensymb}
\usepackage{amsmath}
\usepackage{adjustbox} 
\usepackage{amsfonts}
\usepackage{graphicx}
\usepackage{array}
\usepackage{multirow}
\usepackage{makecell}
\usepackage{enumitem}
\usepackage{upgreek}
\usepackage{graphicx}
\usepackage{epstopdf}
\usepackage{subcaption}
\usepackage{comment}

\usepackage{soul}

\usepackage{algorithm}
\usepackage[noend]{algpseudocode}

\usetikzlibrary{matrix,shapes,arrows,positioning,chains}
\tikzstyle{process} = [rectangle, minimum width=3cm, minimum height=1cm, text centered, draw=black]
\tikzstyle{arrow} = [thick,->,>=stealth]
\tikzstyle{io} = [trapezium, trapezium left angle=70, trapezium right angle=110, text centered]

\begin{document}

\title{Specialized Change Detection using Segment Anything}

\author{Tahir Ahmad and Sudipan Saha  
\thanks{Tahir Ahmad is with Center for Cybersecurity, Fondazione Bruno Kessler, Trento, Italy.E-mail: ahmad@fbk.eu}
\thanks{Sudipan Saha is with Yardi School of Artificial Intelligence, Indian Institute of Technology Delhi, New Delhi, India. E-mail: sudipan.saha@scai.iitd.ac.in}
}

\maketitle

\begin{abstract}
Change detection (CD) is a fundamental task in Earth observation. While most change detection methods detect all changes, there is a growing need for specialized methods targeting specific changes relevant to particular applications while discarding the other changes. For instance, urban management might prioritize detecting the disappearance of buildings due to natural disasters or other reasons. Furthermore, while most supervised change detection methods require large-scale training datasets, in many applications only one or two training examples might be available instead of large datasets. Addressing such needs, we propose a focused CD approach using the Segment Anything Model (SAM), a versatile vision foundation model. Our method leverages a binary mask of the object of interest in pre-change images to detect their disappearance in post-change images. By using SAM's robust segmentation capabilities, we create prompts from the pre-change mask, use those prompts to segment the post-change image, and identify missing objects. This unsupervised approach demonstrated for building disappearance detection, is adaptable to various domains requiring specialized CD. Our contributions include defining a novel CD problem, proposing a method using SAM, and demonstrating its effectiveness. The proposed method also has benefits related to privacy preservation.
\end{abstract}


\IEEEpeerreviewmaketitle

\fboxsep=0mm
\fboxrule=0.1pt

\section{Introduction}
Change detection (CD) plays an important role in Earth observation, playing an important role across various applications. CD involves identifying changes in the state of an area over time, leveraging data from satellite or aerial images. With the emergence of deep learning, most CD methods currently use deep learning-based techniques \cite{saha2019unsupervised}. However, despite significant advancement in CD techniques, most existing CD methods typically identify all changes, which, while useful, may not always be the most efficient approach for specific applications. There is often a need for more specialized change detection methods tailored to particular types of changes relevant to specific fields \cite{hermann2023filtering}. For instance, in the context of urban management, stakeholders might be particularly interested in detecting the disappearance of buildings, which could indicate destruction due to natural disasters like earthquakes. Similarly, in forestry, detecting specific changes such as deforestation might be more relevant than identifying all possible changes in the landscape. This requires a more focused approach to change detection, one that can target specific changes of interest. Additionally, most supervised change detection methods need large-scale datasets for training \cite{saha2022supervised}, which may not be available for all applications.
\par
To address the above-mentioned needs, we can consider a scenario where a binary mask indicating the object of interest in the pre-change image is available. The objective in this scenario is to detect the disappearance of these objects in the post-change image. This approach is useful in several practical applications. For example, in the aftermath of an earthquake, rapid assessment of building damage is important for emergency response and recovery efforts. Traditional change detection methods might identify a variety of changes in the urban landscape, but a specialized method would directly highlight the buildings that have disappeared or been significantly damaged. Furthermore, this approach exploits the pre-disaster building mask which is generally available, whereas a generic change detection method would have discarded this information. 
\par
\begin{figure}[!ht]
	\centering	
		\includegraphics[width=0.5\textwidth]{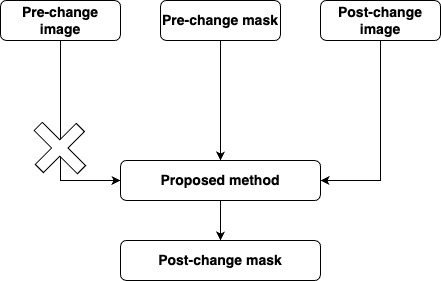}
	\caption{Proposed data scenario: only pre-change mask is used by the proposed method and pre-change image itself is not required.}
\label{figureDataScenario}
\end{figure}

Towards this specific type of change detection, we propose leveraging a recently popularized vision foundation model known as Segment Anything. Vision foundation models have gained popularity due to their versatility and effectiveness in various computer vision tasks. Segment Anything model (SAM), in particular, offers robust segmentation capabilities that can be cleverly adapted for the specialized task of detecting the disappearance of objects.
SAM is designed to segment objects in an image based on prompts, which can be points, bounding boxes, or masks. We cleverly use the pre-change data to create prompts guided by the provided binary mask, ensuring that the focus remains on the relevant objects. The generated prompts are then used as input to segment the post-change image to determine if these segmented objects are still present or if they have disappeared. This approach allows for a more targeted and effective change detection process. Furthermore, the proposed approach does not use any training data and does not train or fine-tune SAM in any manner. Thus, the approach can be considered unsupervised, though we assume the availability of a pre-change object mask. Remarkably, the pre-change image itself is also not required for our approach and only a pre-change mask is required. The data scenario for this task is shown in Figure \ref{figureDataScenario}. Such a data scenario can also be useful for privacy preservation if the pre-change image is held by a different agency than the post-change one. Though, we demonstrate our work in the context of building disappearance detection, the adaptability of SAM implies that this approach can be extended to various other domains requiring specialized change detection. Our contributions are:
\begin{enumerate}
    \item We discuss a novel change detection problem scenario that involves a post-change image and pre-change object mask as input. Furthermore, this problem scenario does not assume the availability of any additional training dataset. This problem scenario may be useful in many applications. 
    \item For change detection in the above-mentioned scenario, we propose a method exploiting the recently popular vision foundation model SAM.
    \item We demonstrate the effectiveness of the proposed approach for building disappearance detection.
\end{enumerate}
The rest of this paper is organized as follows. We discuss some related works in Section \ref{sectionRelatedWorks}. The proposed method is detailed in Section \ref{sectionMethod}. Experimental validation is detailed in Section \ref{sectionExperiments}. Finally, the chapter is concluded in Section \ref{sectionConclusion}.

\section{Related works}
\label{sectionRelatedWorks}
\subsection{Semantic segmentation} Popular semantic segmentation architectures in computer vision include fully convolutional networks (FCNs) \cite{long2015fully}, U-Net \cite{ronneberger2015u}, SegNet \cite{badrinarayanan2017segnet}, and DeepLab \cite{chen2017rethinking}. In addition to computer vision, these architectures are generally effective for Earth observation as well. Nonetheless, these supervised approaches need large amounts of training data for effective supervised learning. Various studies have attempted to train semantic segmentation models with limited or no labeled data, although these generally achieve lower performance compared to fully supervised models. For instance, Hua et al. \cite{hua2021semantic} introduced a method for training semantic segmentation models using sparse annotations. Additionally, Saha et al. \cite{saha2022unsupervised} proposed an unsupervised semantic segmentation model utilizing contrastive learning. 

\subsection{Change detection} Among the various tasks in remote sensing, change detection stands out as one of the most widely studied areas. Change detection involves comparing satellite or aerial images taken at different times to identify significant alterations in the landscape. Its popularity is due to its applications in monitoring environmental changes \cite{mucher2000land}, urban development \cite{saha2020building}, disaster management \cite{saha2020building}, and forestry \cite{hayes2001comparison}.  Change detection methods can be supervised \cite{shi2021deeply}, semisupervised \cite{zhang2018coarse}, and unsupervised \cite{saha2020building} depending on the availability of training data. Nevertheless, unsupervised change detection methods are often preferred due to the challenges in collecting labeled bi-temporal data. Most change detection methods use architectures similar to semantic segmentation \cite{saha2022supervised}, exploiting the similarity between these two tasks. While most change detection works merely focus on identifying changes irrespective of their type, a few works attempt to either identify different types of change \cite{saha2019unsupervised} or focus on filtering specific types of changes \cite{hermann2023filtering} Our work is more related to  \cite{hermann2023filtering} as we are also interested in detecting specific type of changes while discarding other.

\subsection{Vision Transformers} Vision Transformers (ViTs) \cite{dosovitskiy2020image} are a recent innovation in computer vision that challenges the long-standing dominance of convolutional neural networks (CNNs). Unlike CNNs, which process images in a grid-like manner, ViTs decompose images into patches, treating them similarly to words in a sentence. These patches are then processed using transformer architectures, similar to those used in natural language processing, to identify the relationships between them. This approach has demonstrated significant potential in capturing global context, achieving state-of-the-art performance in image recognition tasks.

\subsection{Foundation models} Vision foundation models represent a significant development in the field of computer vision \cite{awais2023foundational}, fundamentally altering the way we analyze and interpret visual data. These models primarily leverage self-supervised learning techniques applied to large datasets, resulting in robust generalization capabilities that span a wide array of applications. Vision Transformers (ViTs) \cite{dosovitskiy2020image} have become an important component of these models due to their effectiveness in capturing long-range dependencies within images. Among the various foundation models, the SAM \cite{kirillov2023segment} stands out for its design tailored to promptable semantic segmentation. This model is particularly relevant to and utilized in our work, as will be detailed later.

\section{Proposed Method}
\label{sectionMethod}
The proposed method works on the following assumptions:
\begin{itemize}
    \item A pre-change binary mask is available indicating all objects belonging to the class of interest. The pre-change image itself may or may not be available and is not used by the proposed method.
    \item The post-change image is available.
    \item The changes of interest are the disappearance of objects.
\end{itemize}
The proposed method uses the Segment Anything model, the functioning of which is outlined in Section \ref{sectionSAM}. Building upon the prompts generated from the pre-change mask the segmentation mask and the confidence output obtained from SAM, the proposed specialized change detection method is described in Section \ref{sectionSpecizedChangeDetection}.

\subsection{Segment Anything}
\label{sectionSAM}
The SAM is designed to efficiently generate segmentation masks for a wide variety of objects in images.  SAM is engineered to be ambiguity-aware, capable of generating multiple valid masks for ambiguous cases. One of SAM's key advantages is its ability to handle diverse prompts, including point, box, and mask prompts, as well as free-form text prompts. This flexibility, combined with its real-time performance and ambiguity-awareness, makes SAM a powerful tool for promptable segmentation tasks. SAM consists of three main components:
\begin{enumerate}
    \item The first component of SAM is the image encoder, which produces an embedding of the input image. This image embedding represents the visual content of the image and serves as input for the subsequent stages of the model.
    \item The prompt encoder is important for SAM, encoding the prompts provided to the model. These prompts can take various forms, including point, box, and mask prompts. By encoding prompts, SAM can incorporate external information to guide the segmentation process.
    \item Finally a decoder ingests the encoder representations and produces a segmentation mask along with a confidence score. For our proposed method, both mask and confidence score are used.
\end{enumerate}
SAM is constructed using Transformer-based vision models, incorporating specific optimizations to ensure real-time performance. By dividing the model into distinct components, SAM facilitates the reuse of image embeddings across various prompts. The separation of architecture into distinct components—image encoder, prompt encoder, and mask decoder—allows each part to specialize and perform its function optimally. 

\begin{figure*}[!ht]
	\centering	
		\includegraphics[width=0.55\textwidth]{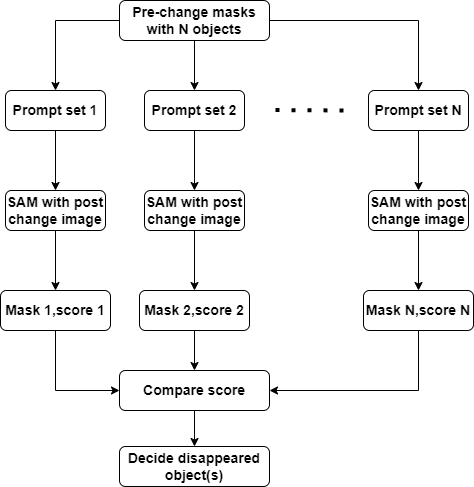}
	\caption{Proposed method}
\label{figureProposedCDMethodWithSAM}
\end{figure*}

\subsection{Specialized change detection}
\label{sectionSpecizedChangeDetection}
Our proposed method is built upon the following concept. Given the object mask corresponding to the pre-change image, we assume there are $N$ total connected objects. For simplicity, let us assume that only one object has changed, meaning this object has been replaced by a different class in the post-change image. To identify this changed object, we can follow a systematic approach using point prompts, which are described as follows:
\begin{enumerate}
    \item The first step in the proposed method is a selection of prompts that will be fed later to SAM. SAM can work with different types of prompts as discussed earlier and in this case, we work with point prompts. Recall that the pre-change object mask is available to us. We can simply select a few points from each object (connected components). However, we would like to study differences in the obtained confidence by choosing or excluding different objects. Thus, we leverage point prompts to systematically exclude each object (identified by the mask) from the post-change image. This exclusion is done during the prompt formulation stage, i.e., we sample a few point prompts from each object, excluding one object at a time. 
    \item We feed the post-change image along with the selected point prompts (excluding a specific object at a time) to the SAM model. SAM generates a segmentation mask and a confidence score based on the prompt and image.
    \item The next step is based on the analysis of the confidence score generated by SAM. When some points from a changed object are included as a prompt, the SAM model will produce a low confidence score since the nature of the changed object in the post-change image differs from the target class in the pre-change image. In other words, a prompt from a changed object can be considered as an erroneous prompt in the context of the post-change image. Conversely, when the changed object is not included as input, the SAM model will yield a higher confidence score.
    \item By analyzing the confidence scores for each exclusion scenario, we can identify the changed object. Specifically, the object that, when excluded, results in a higher confidence score from the SAM model, is the changed object. The proposed method is illustrated in Figure \ref{figureProposedCDMethodWithSAM}.
\end{enumerate}
This method can be extended to identify multiple changed objects by iteratively applying the same logic and adjusting the sampling process to account for multiple exclusions and evaluations. This approach leverages the confidence scoring mechanism of the SAM model to detect discrepancies between pre-change and post-change images, allowing for efficient and accurate identification of changed objects.
\par
We note that our proposed change detection method eliminates the need for a pre-change image. Instead, it utilizes a pre-change mask, significantly enhancing its applicability and convenience in various scenarios. This is particularly advantageous in multi-sensor environments where pre-change and post-change images are captured by different sensors. The reliance on a pre-change mask rather than the actual pre-change image allows integration across diverse sensor types, overcoming compatibility issues typically encountered in multi-sensor setups. Moreover, our method addresses privacy concerns. In cases where the pre-change image is owned by a different organization than the one capturing the post-change image, sharing original images can pose privacy risks. By using a pre-change mask, our approach ensures that other information within the original pre-change image remains private, as there is no need to transfer or disclose the actual image. This aspect encourages collaboration between organizations that might otherwise be reluctant to share data.

\section{Experimental Validation}
\label{sectionExperiments}
\subsection{Data} We evaluated the proposed method for building destruction detection on several scenes from the SECOND dataset \cite{yang2010semantic}.

\begin{figure}[!h]
\centering
\subfigure[]{%
            
         \fbox{\includegraphics[height=2 cm]{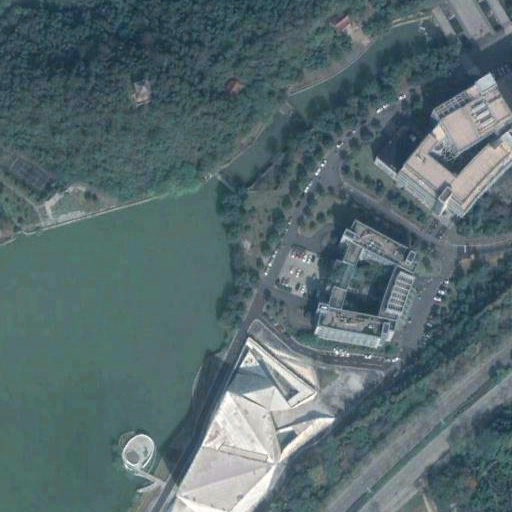}}
            \label{figurePrechange00402}
        }%
\hspace{0.2 cm}
\subfigure[]{%
            
         \fbox{\includegraphics[height=2 cm]{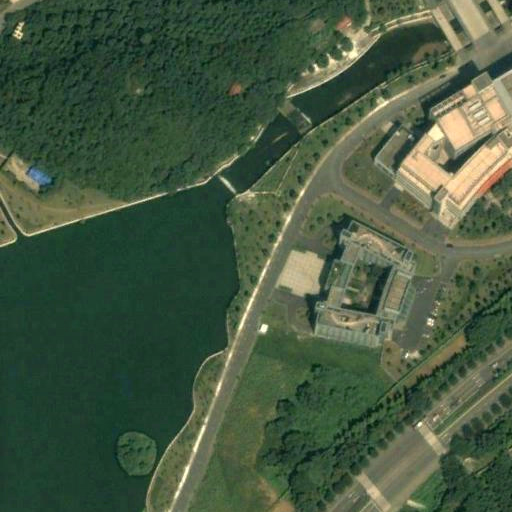}}
            \label{figurePostchange00402}
        }%

\subfigure[]{%
            
         \fbox{\includegraphics[height=2 cm]{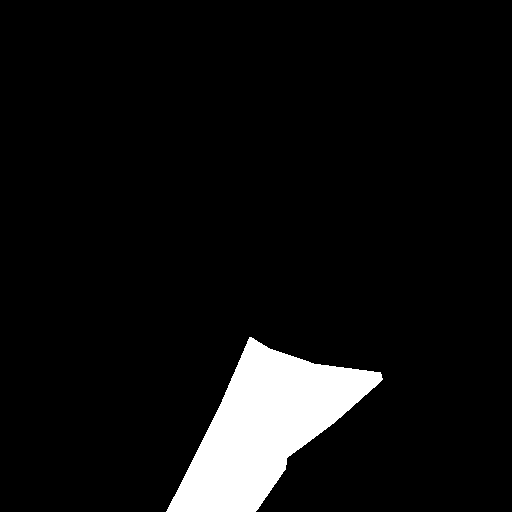}}
            \label{figureReference00402}
        }%
\hspace{0.2 cm}
\subfigure[]{%
            
         \fbox{\includegraphics[height=2 cm]{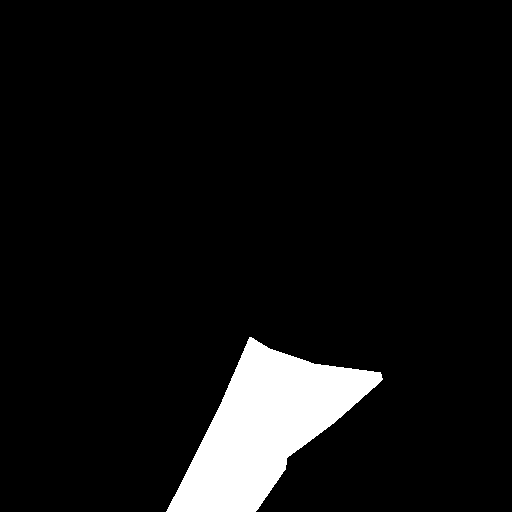}}
            \label{figureResult00402}
        }%

\subfigure[]{%
            
         \fbox{\includegraphics[height=2 cm]{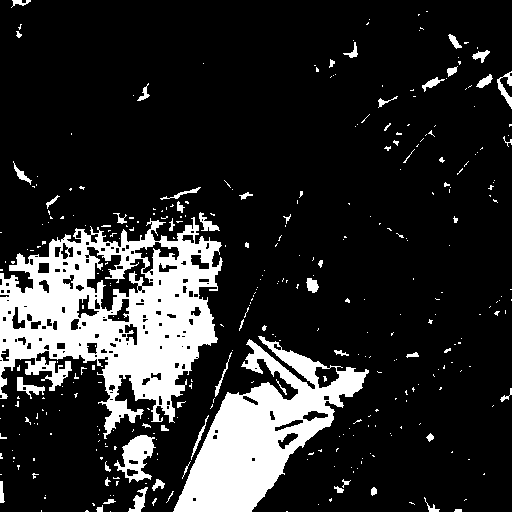}}
            \label{figureRCVA00402}
        }%
\hspace{0.2 cm}
\subfigure[]{%
            
         \fbox{\includegraphics[height=2 cm]{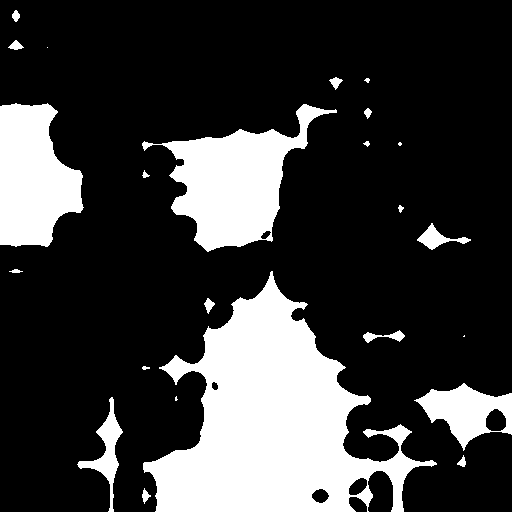}}
            \label{figureDCVA00402}
        }%

\caption{Result visualization on a scene with a single connected changed object: (a) Pre-change image, (b) Post-change image, (c)  Reference, (d) Result obtained by the proposed method, (e) Result obtained by RCVA, (f) Result obtained by DCVA. Changed pixels are shown in white.}
\label{figureResult00402}
\end{figure}

\begin{figure}[!h]
\centering
\subfigure[]{%
            
         \fbox{\includegraphics[height=2 cm]{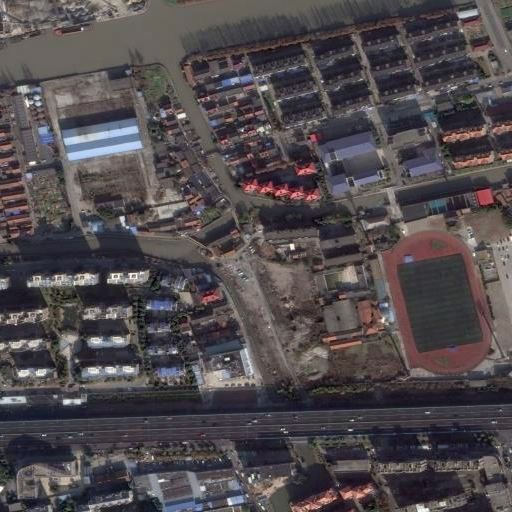}}
            \label{figurePrechange00122}
        }%
\hspace{0.2 cm}
\subfigure[]{%
            
         \fbox{\includegraphics[height=2 cm]{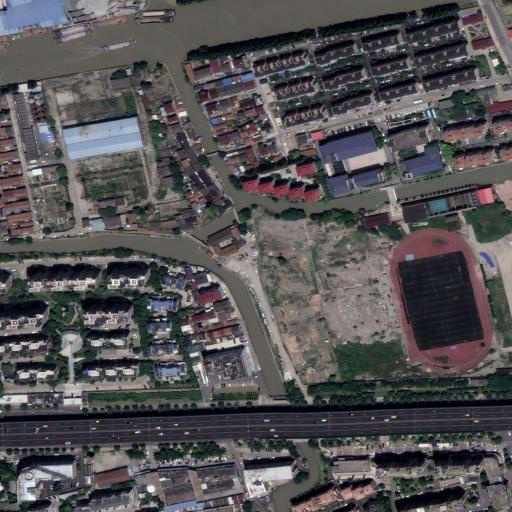}}
            \label{figurePostchange00122}
        }%

\subfigure[]{%
            
         \fbox{\includegraphics[height=2 cm]{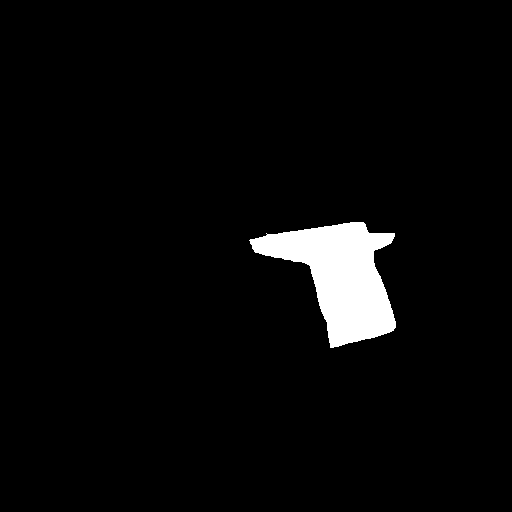}}
            \label{figureReference00122}
        }%
\hspace{0.2 cm}
\subfigure[]{%
            
         \fbox{\includegraphics[height=2 cm]{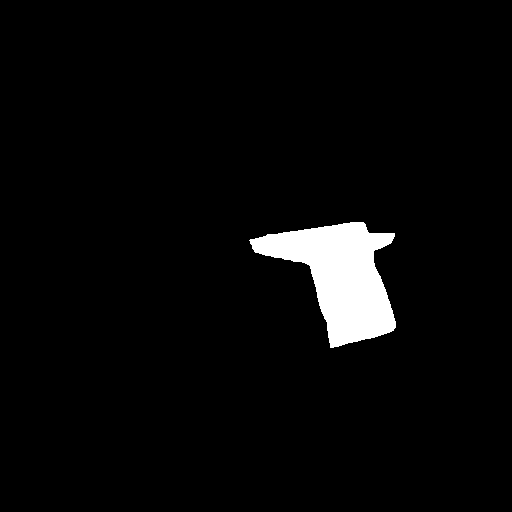}}
            \label{figureResult00122}
        }%

\caption{Another result visualization on a scene with a single connected changed object: (a) Pre-change image, (b) Post-change image, (c)  Reference, (d) Result obtained by the proposed method.}
\label{figureResult00122}
\end{figure}

\subsection{Result for single disappeared object} The results for a complex scene featuring diverse elements such as vegetation, roads, water bodies, and multiple buildings are presented in Figure \ref{figureResult00402}. The figure highlights a significant transformation in vegetation cover between the pre-change and post-change images. Additionally, it reveals that a single, previously connected building has vanished. This scenario shows the difficulty of identifying a single changed building while disregarding other irrelevant changes in the environment.
The proposed method demonstrates its capability to accurately detect this altered building, as depicted in the Figure. In contrast, traditional methods such as Robust Change Vector Analysis (RCVA) \cite{thonfeld2016robust} (Figure \ref{figureRCVA00402})  and Deep Change Vector Analysis \cite{saha2019unsupervised} (Figure \ref{figureDCVA00402})
struggle to deliver satisfactory performance in such complex scenarios. This comparison illustrates the effectiveness of the proposed approach in isolating specific changes of interest along with a backdrop of various other changes. Another similar result is illustrated in Figure \ref{figureResult00122}.

\begin{figure}[!h]
\centering
\subfigure[]{%
            
         \fbox{\includegraphics[height=2 cm]{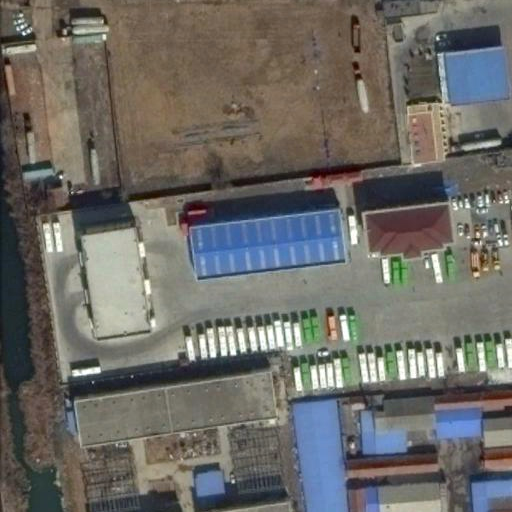}}
            \label{figurePrechange02699}
        }%
\hspace{0.2 cm}
\subfigure[]{%
            
         \fbox{\includegraphics[height=2 cm]{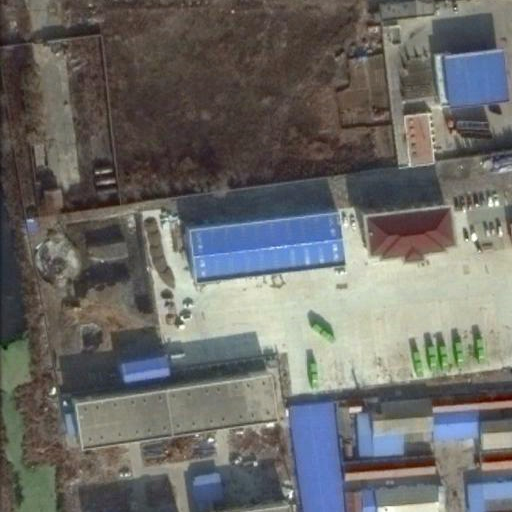}}
            \label{figurePostchange02699}
        }%

\subfigure[]{%
            
         \fbox{\includegraphics[height=2 cm]{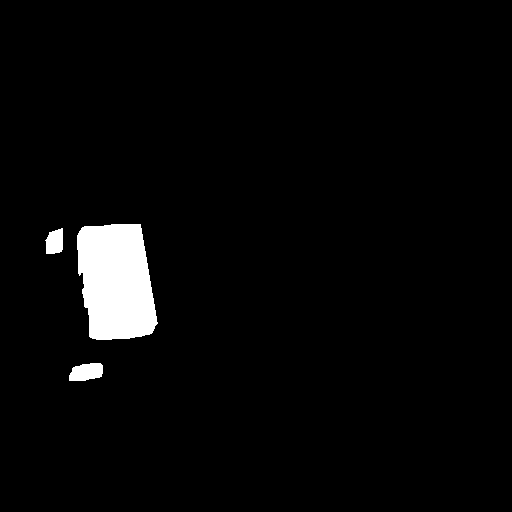}}
            \label{figureReference02699}
        }%
\hspace{0.2 cm}
\subfigure[]{%
            
         \fbox{\includegraphics[height=2 cm]{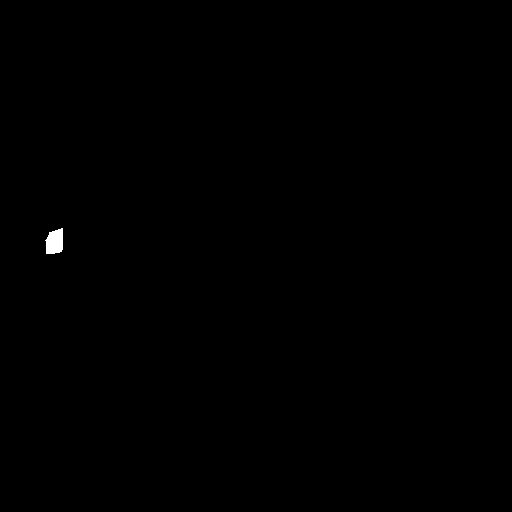}}
            \label{figureResult02699SingleObject}
        }%
\hspace{0.2 cm}
\subfigure[]{%
            
         \fbox{\includegraphics[height=2 cm]{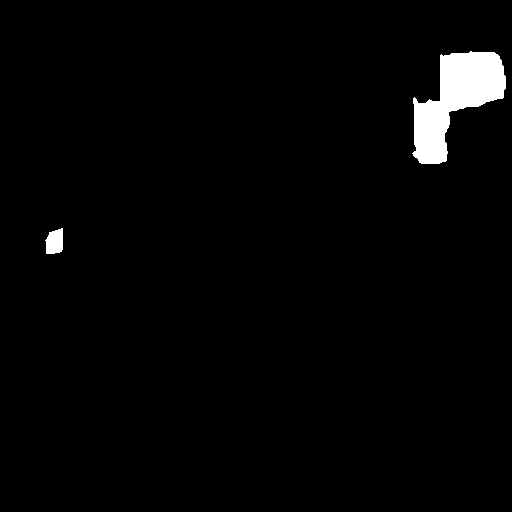}}
            \label{figureResult02699TwoObjects}
        }%

\caption{Result visualization on a scene with a multiple connected changed object: (a) Pre-change image, (b) Post-change image, (c)  Reference, (d) Result obtained by the proposed method considering single changed object, (e) Result obtained by the proposed method considering two changed objects.}
\label{figureResult02699}
\end{figure}

\subsection{Result for multiple disappeared objects} The results for a complex scene containing multiple changed building objects are depicted in Figure \ref{figureResult02699}.  It showcases the capability of the proposed method to detect and highlight these changes accurately, even in the presence of numerous structures undergoing modifications. When the result is obtained with the assumption that there is only one changed object, the proposed method detects one of the changed buildings, as shown in Figure \ref{figureResult02699SingleObject}. However, when the result is obtained with the assumption that there are two changed objects, the proposed method incorrectly detects one unchanged building as changed, as shown in Figure \ref{figureResult02699TwoObjects}. This scenario highlights a limitation of the proposed method: its performance declines in complex scenes with multiple buildings or objects of interest that have changed. This misidentification shows the challenge of accurately detecting changes in such complex environments, where the presence of several potential changes can lead to false positives.

\section{Conclusion}
\label{sectionConclusion}
Foundation models have recently become popular in deep learning and in computer vision. SAM is one such foundation model that has gained popularity due to its performance in semantic segmentation. In this work, we have shown the potential of SAM in a particular data scenario of change detection, which is applicable in several scenarios including for detecting building destruction post-disaster. Through experimental validation, we have shown that the proposed method works reasonably in the proposed setting. A significant advantage of our approach is that it does not involve a training phase, thereby eliminating the need for additional training data and computational resources. This characteristic makes the method more efficient and accessible for practical applications.
It is important to acknowledge, however, that the use of foundation models like SAM in Earth observation tasks, especially change detection, remains in the exploratory stages. While our study demonstrates the promising potential of such models, there is considerable room for further research and development to fully exploit their capabilities.

\bibliographystyle{ieeetr}
\bibliography{sigproc}

\begin{thebibliography}{10}

\bibitem{saha2019unsupervised}
S.~Saha, F.~Bovolo, and L.~Bruzzone, ``Unsupervised deep change vector analysis
  for multiple-change detection in vhr images,'' {\em IEEE Transactions on
  Geoscience and Remote Sensing}, vol.~57, no.~6, pp.~3677--3693, 2019.

\bibitem{hermann2023filtering}
M.~Hermann, S.~Saha, and X.~X. Zhu, ``Filtering specialized change in a
  few-shot setting,'' {\em IEEE Journal of Selected Topics in Applied Earth
  Observations and Remote Sensing}, vol.~16, pp.~1185--1196, 2023.

\bibitem{saha2022supervised}
S.~Saha, M.~Shahzad, P.~Ebel, and X.~X. Zhu, ``Supervised change detection
  using prechange optical-sar and postchange sar data,'' {\em IEEE Journal of
  Selected Topics in Applied Earth Observations and Remote Sensing}, vol.~15,
  pp.~8170--8178, 2022.

\bibitem{long2015fully}
J.~Long, E.~Shelhamer, and T.~Darrell, ``Fully convolutional networks for
  semantic segmentation,'' in {\em Proceedings of the IEEE Conference on
  Computer Vision and Pattern Recognition}, pp.~3431--3440, 2015.

\bibitem{ronneberger2015u}
O.~Ronneberger, P.~Fischer, and T.~Brox, ``U-net: Convolutional networks for
  biomedical image segmentation,'' in {\em International Conference on Medical
  image computing and computer-assisted intervention}, pp.~234--241, Springer,
  2015.

\bibitem{badrinarayanan2017segnet}
V.~Badrinarayanan, A.~Kendall, and R.~Cipolla, ``Segnet: A deep convolutional
  encoder-decoder architecture for image segmentation,'' {\em IEEE Transactions
  on Pattern Analysis and Machine Intelligence}, vol.~39, no.~12,
  pp.~2481--2495, 2017.

\bibitem{chen2017rethinking}
L.-C. Chen, G.~Papandreou, F.~Schroff, and H.~Adam, ``Rethinking atrous
  convolution for semantic image segmentation,'' {\em arXiv preprint
  arXiv:1706.05587}, 2017.

\bibitem{hua2021semantic}
Y.~Hua, D.~Marcos, L.~Mou, X.~X. Zhu, and D.~Tuia, ``Semantic segmentation of
  remote sensing images with sparse annotations,'' {\em arXiv preprint
  arXiv:2101.03492}, 2021.

\bibitem{saha2022unsupervised}
S.~Saha, M.~Shahzad, L.~Mou, Q.~Song, and X.~X. Zhu, ``Unsupervised
  single-scene semantic segmentation for earth observation,'' {\em IEEE
  Transactions on Geoscience and Remote Sensing}, vol.~60, pp.~1--11, 2022.

\bibitem{mucher2000land}
C.~Mucher, K.~Steinnocher, F.~Kressler, and C.~Heunks, ``Land cover
  characterization and change detection for environmental monitoring of
  pan-europe,'' {\em International Journal of Remote Sensing}, vol.~21,
  no.~6-7, pp.~1159--1181, 2000.

\bibitem{saha2020building}
S.~Saha, F.~Bovolo, and L.~Bruzzone, ``Building change detection in {VHR SAR}
  images via unsupervised deep transcoding,'' {\em IEEE Transactions on
  Geoscience and Remote Sensing}, vol.~59, no.~3, pp.~1917--1929, 2020.

\bibitem{hayes2001comparison}
D.~J. Hayes and S.~A. Sader, ``Comparison of change-detection techniques for
  monitoring tropical forest clearing and vegetation regrowth in a time
  series,'' {\em Photogrammetric engineering and remote sensing}, vol.~67,
  no.~9, pp.~1067--1075, 2001.

\bibitem{shi2021deeply}
Q.~Shi, M.~Liu, S.~Li, X.~Liu, F.~Wang, and L.~Zhang, ``A deeply supervised
  attention metric-based network and an open aerial image dataset for remote
  sensing change detection,'' {\em IEEE transactions on geoscience and remote
  sensing}, vol.~60, pp.~1--16, 2021.

\bibitem{zhang2018coarse}
W.~Zhang, X.~Lu, and X.~Li, ``A coarse-to-fine semi-supervised change detection
  for multispectral images,'' {\em IEEE Transactions on Geoscience and Remote
  Sensing}, vol.~56, no.~6, pp.~3587--3599, 2018.

\bibitem{dosovitskiy2020image}
A.~Dosovitskiy, L.~Beyer, A.~Kolesnikov, D.~Weissenborn, X.~Zhai,
  T.~Unterthiner, M.~Dehghani, M.~Minderer, G.~Heigold, S.~Gelly, {\em et~al.},
  ``An image is worth 16x16 words: Transformers for image recognition at
  scale,'' {\em arXiv preprint arXiv:2010.11929}, 2020.

\bibitem{awais2023foundational}
M.~Awais, M.~Naseer, S.~Khan, R.~M. Anwer, H.~Cholakkal, M.~Shah, M.-H. Yang,
  and F.~S. Khan, ``Foundational models defining a new era in vision: A survey
  and outlook,'' {\em arXiv preprint arXiv:2307.13721}, 2023.

\bibitem{kirillov2023segment}
A.~Kirillov, E.~Mintun, N.~Ravi, H.~Mao, C.~Rolland, L.~Gustafson, T.~Xiao,
  S.~Whitehead, A.~C. Berg, W.-Y. Lo, {\em et~al.}, ``Segment anything,'' in
  {\em Proceedings of the IEEE/CVF International Conference on Computer
  Vision}, pp.~4015--4026, 2023.

\bibitem{yang2010semantic}
K.~Yang, G.~Xia, Z.~Liu, B.~Du, W.~Yang, M.~Pelillo, and L.~Zhang, ``Semantic
  change detection with asymmetric siamese networks. arxiv 2020,'' {\em arXiv
  preprint arXiv:2010.05687}.

\bibitem{thonfeld2016robust}
F.~Thonfeld, H.~Feilhauer, M.~Braun, and G.~Menz, ``Robust change vector
  analysis ({RCVA}) for multi-sensor very high resolution optical satellite
  data,'' {\em International Journal of Applied Earth Observation and
  Geoinformation}, vol.~50, pp.~131--140, 2016.

\end{thebibliography}

\end{document}